\documentclass[pra,twocolumn,amsmath,amssymb,floatfix,reprint,footinbib,superscriptaddress,longbibliography,showkeys]{revtex4-2}
\usepackage{picinpar,graphicx}

\usepackage{braket}
\usepackage{amsmath}
\usepackage{graphicx}
\usepackage{epstopdf}
\usepackage{dcolumn}
\usepackage{graphicx}
\usepackage{mathrsfs}
\usepackage{mdwlist}
\usepackage{subfigure}
\usepackage{booktabs}
\usepackage{amsmath}
\usepackage{dsfont}
\usepackage{amstext}
\usepackage{amssymb}
\usepackage{amsbsy}
\usepackage{bbm}
\usepackage{amsthm}
\usepackage{graphicx}
\usepackage{textcomp}
\usepackage{color}
\usepackage[colorlinks,citecolor=blue]{hyperref}
\usepackage{lipsum}
\definecolor{Dgreen}{RGB}{0, 100, 0}
\usepackage{url}
\usepackage[colorlinks]{hyperref}
\hypersetup{%
	plainpages=true,
	breaklinks=true,  
	hypertexnames=false,  
	pageanchor=true,
	colorlinks=true,
	linkcolor={blue},
	citecolor={red},
	urlcolor={blue},
	anchorcolor={black}
}

\begin{document}
	
	\title{Dissecting the superradiant phase transition in the anisotropic Rabi model: Pattern competition and cavity-QED simulation}
	\author{Yuan Qiu}
	\affiliation{Fujian Key Laboratory of Quantum Information and Quantum Optics, Fuzhou University, Fuzhou 350116, China}
	\affiliation{Department of Physics, Fuzhou University, Fuzhou, 350116, China}

	\author{Ke-Xiong Yan}
	\affiliation{Fujian Key Laboratory of Quantum Information and Quantum Optics, Fuzhou University, Fuzhou 350116, China}
	\affiliation{Department of Physics, Fuzhou University, Fuzhou, 350116, China}

	\author{Jun-Hao Lin}
	\affiliation{Fujian Key Laboratory of Quantum Information and Quantum Optics, Fuzhou University, Fuzhou 350116, China}
	\affiliation{Department of Physics, Fuzhou University, Fuzhou, 350116, China}

	\author{Jie Song}
	\affiliation{Department of Physics, Harbin Institute of Technology, Harbin 150001, China}
	
	\author{Ye-Hong Chen}\thanks{yehong.chen@fzu.edu.cn}
	\affiliation{Fujian Key Laboratory of Quantum Information and Quantum Optics, Fuzhou University, Fuzhou 350116, China}
	\affiliation{Department of Physics, Fuzhou University, Fuzhou, 350116, China}
	\affiliation{Quantum Information Physics Theory Research Team, Center for Quantum Computing, RIKEN, Wako-shi, Saitama 351-0198, Japan}
	
	\author{Yan-Xia}\thanks{xia-208@163.com}
	\affiliation{Fujian Key Laboratory of Quantum Information and Quantum Optics, Fuzhou University, Fuzhou 350116, China}
	\affiliation{Department of Physics, Fuzhou University, Fuzhou, 350116, China}

	\date{\today}
	
\begin{abstract}\label{key}	
	In this manuscript, we analyze the mechanism of the superradiant phase transition in the anisotropic Rabi model under the classical oscillator limit using the pattern picture. By expanding the anisotropic Rabi model Hamiltonian in operator space, we obtained three patterns, and we find that the phase transition arises from the competition between patterns. The difficulty in achieving the classical oscillator limit motivates our investigation into the quantum phase transition within a parametrically-driven Jaynes-Cummings model. This parametrically-driven Jaynes-Cummings model can reproduce the dynamics of an ultrastrong-coupling anisotropic Rabi model in a squeezed-light frame. According to the eigenenergies and eigenstates of the normal and superradiant phases of this equivalent anisotropic Rabi model, we find that the excitation energy of the normal phase and the superradiant phase vanishes at the critical point. The photon number becomes infinite beyond the critical point. These results indicate that the system undergoes a superradiant phase transition at the critical point. 
\end{abstract} 
\maketitle

\label{key} 
\section{INTRODUCTION}
Since 1937, Landau and Ginzburg proposed a universal method for dealing with phase transition by introducing the concept of order parameter, the study of modern phase transitions and critical behavior has formed a mature theoretical framework~\cite{Landau:480039,carr2010understanding}. When the coupling strength reaches a critical value, phenomena such as macroscopic occupation of photons and abrupt changes in the system's ground state indicate the occurrence of quantum phase transition (QPT)~\cite{PhysRevLett.115.180404,PhysRevA.95.013819,PhysRevLett.117.123602,PhysRevLett.131.113601}. QPT has become one of the core topics in condensed matter physics~\cite{PhysRevB.65.140405,kleeorin2018quantum,PhysRevB.106.014313} and quantum optics~\cite{Zhang_2020,PhysRevLett.124.073602}, due to its characteristic of being dominated by quantum fluctuations at absolute zero, and has been widely studied in various light-matter coupling systems, such as the Dicke-type system~\cite{emary2003chaos,Dimer2006ProposedRO,PhysRevA.7.831,PhysRevLett.108.043003,PhysRevB.75.024402,PhysRevLett.104.130401,PhysRevA.19.2392,PhysRevA.85.043821,PhysRevLett.92.073602}. The Dicke-type system describes the interaction between a single-mode boson field and an ensemble of $N$ two-level atoms, which exhibits a superradiant phase transition at thermodynamic limits, i.e., $N\rightarrow \infty$~\cite{Emary2003,PhysRevLett.90.044101}. 

However, reaching the thermodynamic limit is one of the main challenges in achieving QPT in laboratories~\cite{PhysRevA.95.013819}. This challenge has prompted researchers to turn their attention to the study of the QPT in finite-sized systems, such as the quantum Rabi model~\cite{PhysRevLett.115.180404,PhysRevA.94.023835,PhysRevA.109.023721,cai2021observation,ying2018quantum,PhysRevA.104.063703,PhysRevA.87.013826}, which is a simplified version of the Dicke model with $N=1$. In this model, the classical oscillator limit (i.e.,~the ratio of atomic transition frequency $\omega_q$ to cavity field frequency $\omega_c$ approaches infinity) replaces the thermodynamic limit and also theoretically predicts the QPT~\cite{chen2021experimental}. The anisotropic Rabi model can be defined as the generalization of the spin-boson Rabi model~\cite{PhysRevX.4.021046}, unlike the standard quantm Rabi model, the anisotropic Rabi model has different coupling strengths for the rotating-wave and counter-rotating-wave interaction terms~\cite{zhang2015analytical}. Previous studies have shown that this model also undergoes a phase transition in the classical oscillator limit~\cite{PhysRevA.104.063703}. However, the physical mechanism responsible for this phase transition has not been thoroughly studied.

In this manuscript, we diagonalize the anisotropic Rabi model Hamiltonian in the operator space, gaining a deeper understanding of the reasons for the occurrence of QPT in the anisotropic Rabi model. The results show that there is a competitive relationship among the three patterns in the superradiant phase transition, which explains why and how the superradiant phase transition occurs. We also discuss the influence of the rotating- to counter-rotating-wave coupling ratio on the pattern competition and the contributions of the patterns to the fidelity out-of-time-order correlation function. Regarding the difficulty of reaching the classical oscillator limit in quantum phase transition and the challenges posed by the no-go theorem~\cite{nataf2010no,chen2021experimental,puebla2013excited,lamberto2024superradiant,rzazewski1975phase}, we apply a parametric amplification method in a generic cavity quantum electrodynamics system to obtain a simulated anisotropic Rabi model to explore the QPT.  Specifically, we apply a two-photon drive in the Jaynes-Cummings (JC) model~\cite{Shore1993,PhysRevA.101.043835,PhysRevLett.120.093602} to obtain a simulated anisotropic Rabi model. The effective cavity frequency and the equivalent coupling of this simulated quantum Rabi model are related to the two-photon drive strength. Therefore, we can control the occurrence of the phase transition by changing the drive strength. 
 
 To investigate the critical behavior of this simulated anisotropic Rabi model, we derive its effective low-energy Hamiltonian in the classical oscillator limit, and obtain the eigenenergies and eigenstates of the normal phase and the superradiant phase. The simulation results indicate that when the coupling strength approaches the critical point, the system exhibits unique critical behaviors that mark the QPT. Specifically, the excitation energy becomes zero. The ground-state energy is continuous at the critical point, while the second-order derivative of the ground-state energy to the coupling strength is discontinuous at the critical point. When the coupling strength exceeds the critical point, the photon number exhibits macroscopic occupation. 

The structure of the manuscript is as follows. In Sec.~\ref{s2} we review the key aspects of Shen et al.'s work ~\cite{PhysRevA.95.013819} on quantum phase transition. In Sec.~\ref{s3} we diagonalize the anisotropic Rabi model in the operator space and analyze why and how the phase transition occurs. In Sec.~\ref{s4} we present the model and Hamiltonian, and derive the
eigenenergies and eigenstates of the normal phase and
the superradiant phase. In Sec.~\ref{s5} we demonstrate that the simulated anisotropic Rabi model undergoes a phase transition at the critical point, and prove that the occurrence of this transition can be controlled by tuning the squeezing parameter. Finally, we conclude our manuscript in Sec.~\ref{s6}.

\section{ PHASE TRANSITION IN THE ANISOTROPIC RABI MODEL: PRIOR WORK}\label{s2}
The anisotropic Rabi model features unequal coupling strengths for its rotating- and counter-rotating-wave interactions. In Ref.~\cite{PhysRevA.95.013819}, Shen et al. investigated the phase transitions of the anisotropic Rabi model under the classical oscillator limit ($\Omega/\omega_0\to\infty$) and the Hamiltonian of the anisotropic Rabi model they studied is denoted as

\begin{eqnarray}
	\begin{aligned}
		H_{\mathrm{AN}}=&\omega_0a^{\dagger}a+\frac{\Omega}{2}\sigma_z-\xi_{\text{R}}(a\sigma_+ +a^{\dagger}\sigma_-)\\&-\xi_{\text{CR}}(a\sigma_- +a^{\dagger}\sigma_+).
	\end{aligned}\label{eq1} 
\end{eqnarray}
Here, $a^{\dagger} (a)$ is a creation (annihilation) opetator for the harmonic oscillator with the frequency $\omega_0$, $\sigma_+ (\sigma_-)$ and $\sigma_z$ are Pauli matrices for the qubit with the transition frequency $\Omega$. $\xi_{\text{R}}$ and $\xi_{\text{CR}}$ characterize the qubit-oscillator coupling strengths of the rotating-wave and counter-rotating-wave interactions, respectively. By projecting the effective Hamiltonians of the normal and superradiant phases onto their corresponding spin-down subspaces, the low-energy effective Hamiltonians for both phases can be obtained.

The low-energy effective Hamiltonian of the normal phase is
\begin{eqnarray}
	\begin{aligned}
		H^{\mathrm{AN}}_{\mathrm{np}}&=(\omega_0-\frac{\xi_{\text{R}}^2+\xi_{\text{CR}}^2}{\Omega})a^{\dagger}a-\frac{\xi_{\text{R}}\xi_{\text{CR}}}{\Omega}(a^{\dagger 2}+a^2)\\&-\frac{\xi_{\text{CR}}^2}{\Omega}-\frac{\Omega}{2}.
	\end{aligned}\label{eq2} 
\end{eqnarray}
According to the squeezing transformation $S_{\mathrm{AN}}^{\dagger}(r_{\mathrm{AN}})H^{\mathrm{AN}}_{\mathrm{np}}S_{\mathrm{AN}}(r_{\mathrm{AN}})$, where $S_{\mathrm{AN}}(r_{\mathrm{AN}})=\mathrm{exp}[r_{\mathrm{AN}}(a^{\dagger 2}-a^2)/2],$ when the coefficient of the $(a^{\dagger 2}+a^2)$, the diagonalized Hamiltoniain of the normal phase is
\begin{eqnarray}
	\begin{aligned}
		H^{\mathrm{AN}'}_{\mathrm{np}}=\epsilon^{\mathrm{AN}}_{\mathrm{np}}a^{\dagger}a+E^{\mathrm{AN}}_{\mathrm{np}},
	\end{aligned}
\end{eqnarray}
with the excitation energy,
\begin{eqnarray}
	\begin{aligned}
		\epsilon^{\mathrm{AN}}_{\mathrm{np}}= \sqrt{(\omega_{0}- \frac{\xi_{\text{R}}^2 + \xi_{\text{CR}}^2}{2})^2-(\frac{2\xi_{\text{R}}\xi_{\text{CR}}}{\Omega})^2},
	\end{aligned}
\end{eqnarray}
and the ground-state energy is
\begin{eqnarray}
	\begin{aligned}
		E^{\mathrm{AN}}_{\mathrm{np}}= \frac{1}{2}(\epsilon^{\mathrm{AN}}_{\mathrm{np}}-\omega_0+\frac{\xi_{\text{R}}^2+\xi_{\text{CR}}^2}{\Omega}-\Omega).
	\end{aligned}
\end{eqnarray}
For $w_0 > (\xi_{\text{R}}^2 + \xi_{\text{CR}}^2)/\Omega$, the excitation energy $\epsilon^{\mathrm{AN}}_{\mathrm{np}}$ is real only when $w_0 - (\xi_{\text{R}}^2 + \xi_{\text{CR}}^2)/\Omega\geq (2\xi_{\text{R}}\xi_{\text{CR}})/\Omega$,
then the critical coupling strength is
\begin{eqnarray}
	\begin{aligned}
	\xi_c=\frac{\xi_{\text{R}}+\xi_{\text{CR}}}{\sqrt{\omega_0\Omega}}\leq1.
	\end{aligned}\label{eq6}
\end{eqnarray}
The eigenstates and eigenenergies of $H^{\mathrm{AN}'}_\mathrm{np}$ are
\begin{eqnarray}
	\ket{\phi_{\mathrm{np}}^n} = S(r_{\mathrm{np}}) \ket{n} \ket{\downarrow},
\end{eqnarray}
\begin{eqnarray}
	E_{\mathrm{np}}^n=n\epsilon_{\mathrm{np}}+E_{\mathrm{np}}.
\end{eqnarray}

For the superradiant phase, its low-energy effective Hamiltonian is
\begin{eqnarray}
	\begin{aligned}
	H^{\mathrm{AN}}_{\mathrm{sp}} \simeq &w_0 a^\dagger a 
	- \frac{(\xi_{\text{R}} + \xi_{\text{CR}})^2 \xi_c^{-4} + (\xi_{\text{R}} - \xi_{\text{CR}})^2}{2\Omega \xi_c^2}a^\dagger a \\
	&- \frac{(\xi_{\text{R}} + \xi_{\text{CR}})^2 \xi_c^{-4} - (\xi_{\text{R}} - \xi_{\text{CR}})^2}{4\Omega \xi_c^2}(a^{\dagger 2} + a^2) \\
	&- \frac{(\xi_{\text{R}} + \xi_{\text{CR}})\xi_c^{-2} - (\xi_{\text{R}} - \xi_{\text{CR}})}{2\Omega \xi_c^2} - \frac{\Omega}{4}\big( \xi_c^2 + \xi_c^{-2}\big),
	\end{aligned}
\end{eqnarray}
with the excitation energy,
\begin{eqnarray}
	\begin{aligned}
		\epsilon^{\mathrm{AN}}_{\mathrm{sp}} = w_0 \frac{2\sqrt{\xi_{\text{R}} \xi_{\text{CR}}}}{\xi_{\text{R}} + \xi_{\text{CR}}} \sqrt{1 - \xi_c^{-4}},
	\end{aligned}
\end{eqnarray}
and the ground-state energy
\begin{eqnarray}
	\begin{aligned}
		E^{\mathrm{AN}}_{\mathrm{sp}}= &\frac{1}{2} \left( \epsilon^{\mathrm{AN}}_{\mathrm{sp}} - w_0 - \frac{\xi_{\text{R}}^2 + \xi_{\text{CR}}^2}{\Omega} \right) - \frac{\Omega}{4} \left( \xi_c^2 + \xi_c^{-2} \right) \\&- \frac{(\xi_{\text{R}} + \xi_{\text{CR}})\xi_c^{-2} - (\xi_{\text{R}} - \xi_{\text{CR}})}{2\Omega \xi_c^2}.
	\end{aligned}
\end{eqnarray}
The eigenstates and eigenenergies of $H^{\mathrm{AN}}_{\mathrm{sp}}$ are written as
\begin{eqnarray}
	\begin{aligned}
		|\phi_{\text{sp}}^n\rangle = D(\pm \alpha_0) S(r_{\text{sp}}) |n\rangle |\downarrow^{\pm}\rangle.
	\end{aligned}
\end{eqnarray}

Reference~\cite{PhysRevA.95.013819} further analyzed the excitation energy, ground-state energy, second-order derivative of the ground-state energy, and photon number in the two-dimensional parameter space $(\xi_{\text{R}},\xi_{\text{CR}})$. The numerical results demonstrate that the excitation energies of both the normal and superradiant phases vanish when $\xi_c\to 1$, the ground-state energy remains continuous at the critical coupling strength while its second derivative exhibits a discontinuity, and the order parameter $\langle a^{\dagger}a\rangle$ jumps from zero to infinity. These conclusively verified the occurrence of a QPT in the anisotropic Rabi model at the critical coupling strength.
The simulation results in a two parameter space uncovered distinctive physical behaviors under the critical conditions of QPT. Specifically, the critical condition in the isotropic Rabi model is a point in a one-dimensional parameter space, but in the anisotropic Rabi model it is a straight line in a two-dimensional parameter space which essent\textbf{}ially extends the dimensionality of the QPT.
\section{Dissecting Phase Transition in the Anisotropic Rabi Model}\label{s3}
\begin{figure*}
	\centering
	\includegraphics[scale=0.8]{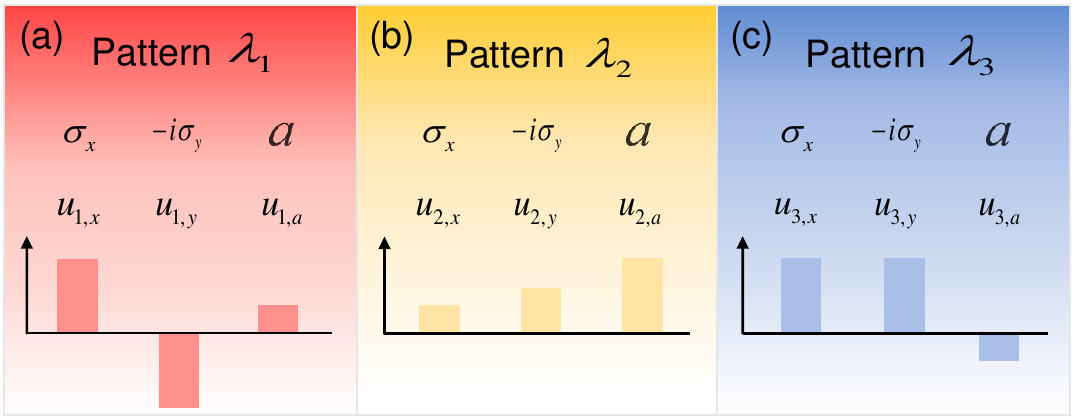}
	\caption{Marks of the patterns obtained by diagonalizing the Hamiltonian $H_{\text{AN}}$. The color bars in the figure represent the approximate weights of $u_{n,j}$ in each pattern, where $n=1,2,3$ and $j=x,y,a$. Here, the free phase factor $e^{i\pi}$ is omitted, the sign of the phase does not affect the fundamental properties of the patterns.}
	\label{F1}
\end{figure*}

 In Sec.~\ref{s2}, we reviewed how Shen et al. successfully characterized the critical behavior of the anisotropic Rabi model using low-energy effective Hamiltonians. However, a microscopic understanding of how competing interactions drive this phase transition remains incomplete. In this section, we diagonalize the Hamiltonian $H_\mathrm{AN}$ in Eq.~$(\ref{eq1})$ in the operator space to obtain three fundamental patterns for dissecting QPT~\cite{Yang_2024,PhysRevB.75.235325}. These three patterns are represented as $\lambda_1$, $\lambda_2$ and $\lambda_3$, which are exact because they can precisely reproduce the Hamiltonian matrix for a given Fock basis.	By using the relation $\sigma_x\sigma_y=i\sigma_z$, the Hamiltonian $H_\mathrm{AN}$ in Eq.~$(\ref{eq1})$ can be reformulated as follows:

\begin{align}
	H_{\mathrm{AN}} &= 
	\begin{pmatrix}
		\sigma_x & i\sigma_y & a^\dagger
	\end{pmatrix}
	M
	\begin{pmatrix}
		\sigma_x \\
		-i\sigma_y \\
		a
	\end{pmatrix}\nonumber\\
	&= \sum_{n=1}^{3} \lambda_n A_n^\dagger A_n, \label{eq25} \\
	A_n &= u_{n,x}(\sigma_x) + u_{n,y}(-i\sigma_y) + u_{n,a}a,\label{eq27}	
\end{align}
where
\begin{equation}
M= \begin{pmatrix}
	0 & \frac{\Omega}{4} & -\frac{\xi_{\text{R}}+\xi_{\text{CR}}}{2}\\
	\frac{\Omega}{4}& 0 & \frac{\xi_{\text{CR}}-\xi_{\text{R}}}{2} \\
	-\frac{\xi_{\text{R}}+\xi_{\text{CR}}}{2} &\frac{\xi_{\text{CR}}-\xi_{\text{R}}}{2} & \omega_0
\end{pmatrix}.
\end{equation}
Here, $\lambda_n$ and $u_n$ $(n=1,2,3)$ are the eigenvalues and eigenvectors of the matrix in Eq.~$(\ref{eq25})$, respectively. They form three basic patterns represented by eigenstates and eigenvectors, as shown in Fig.~$\ref{F1}$. In this pattern picture, the commutation relation of the creation and annihilation operators $[\lambda_nA^{\dagger}aA,\lambda_nA^{\dagger}a^{\dagger}A]\neq 1$. This is because $A_n$ is not a unitary operator and 
	$[A_n,A^{\dagger}_n]\neq 1$, it is a mixture of the boson operator ($a$) and the atomic spin operators ($\sigma_x$ and $-i\sigma_y$). The “projection” here is not the mathematical orthogonal projection, but rather to the action of the pattern operators, which project $a$ and $a^{\dagger}$ into the corresponding patterns~\cite{Yang2023}. The fact that the pattern operator $A_n$ is non-unitary is precisely what gives physical meaning to our pattern decomposition: altering one pattern inevitably influences the others ~\cite{Yang2025}. This competition becomes crucial near the phase transition point, where the redistribution of energy among the patterns drives the system into the superradiant phase (see below). Before discussing the phase transition with pattern picture, we will first verify the validity of the pattern picture.   

\subsection{Patterns and solution}
Equation~$(\ref{eq25})$ can be solved by expanding the system states in the complete basis $|\Gamma_z,m\rangle$, where
\begin{align}
	\sigma_z|\Gamma_z\rangle=\pm(\uparrow,\downarrow)|\Gamma_z\rangle,
\end{align}
 denotes the spin eigenstate along the $z$-direction and  $a^\dagger a|m\rangle=m|m\rangle$ $(m=0,1,2,...,N)$ denotes the truncated Fock basis with the photon number $m$. The matrix elements of $A_n$ in this basis are constructed as
\begin{align}
	[A_n]_{\Gamma_z,m; \Gamma'_z,m'}=\langle\Gamma_z,m|A_n|m',\Gamma'_z\rangle.
\end{align}
Subsequently, Eq.~$(\ref{eq25})$ can be addressed by diagonalizing the matrix as follows
\begin{align}
	&\langle\Gamma_z,m|H_{\mathrm{AN}}|m',\Gamma'_z\rangle=\sum_{n=1}^{3} \lambda_n \langle \Gamma_z, m | A_n^\dagger A_n | m', \Gamma_z' \rangle \nonumber \\
	&= \sum_{n=1}^{3} \lambda_n \sum_{\Gamma_z'', m''} [A_n^\dagger]_{m, \Gamma_z; m'', \Gamma_z''} [A_n]_{m'', \Gamma_z''; m', \Gamma_z'}.\label{eq28}
\end{align}
Upon obtaining the eigenstate wavefunctions $\Psi_i$ (where $i=0,1,...$, correspondings to the ground state, the excited state, and so on), these wavefunctions are projected onto different patterns to determine the contributions of these patterns to the target physical quantities. Taking energy as an example, the energy contributions of the three patterns to the eigenstate $|\Psi_i\rangle$ can be calculated by the following equation
\begin{align}
	E_{\lambda_n}= \langle\Psi_i |\lambda_nA_n^\dagger A_n |\Psi_i \rangle,\, ~(n= 1, 2, 3)
\end{align}
By substituting $A_n$ in Eq.~$(\ref{eq25})$ into $\langle\Psi_i |A_n^\dagger A_n |\Psi_i \rangle$, the relevant physical observable can be expressed as $\langle\Psi_i |a^\dagger a |\Psi_i \rangle_{\lambda_n}=u_{n,a}^2\langle\Psi_i |a^\dagger a |\Psi_i \rangle$ and  $\langle\Psi_i |\sigma_z |\Psi_i \rangle_{\lambda_n}=u_{n,x}u_{n,y}\langle\Psi_i |\sigma_z |\Psi_i \rangle$, which can be easily calculated after diagonalizing the Hamiltonian matrix as given in Eq.~$(\ref{eq28})$ to obtain $|\Psi_i\rangle$. Depending on the chosen operators for decomposing the
Hamiltonian, different observables can be obtained.

\begin{figure}[htbp]
 	\includegraphics[scale=0.583]{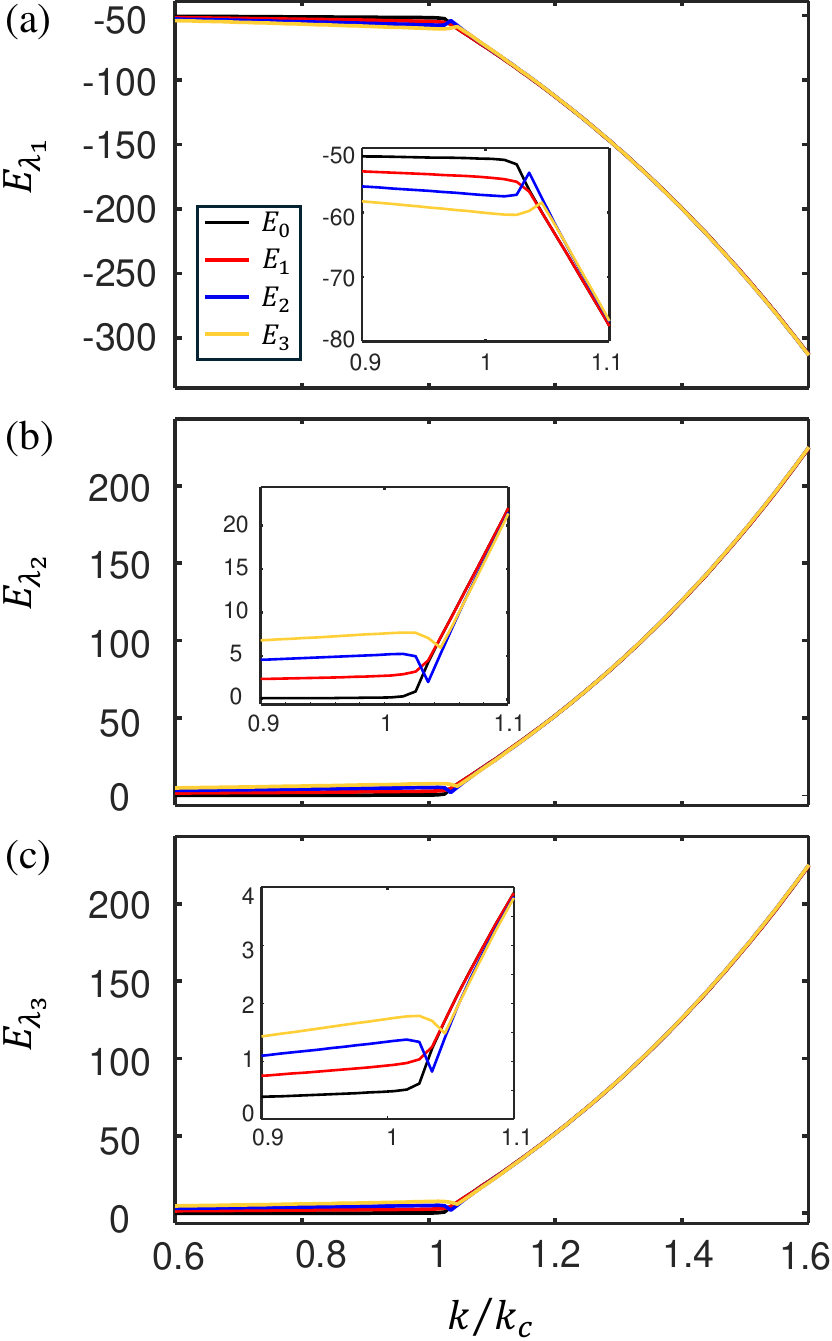}
 	\caption{ First four pattern energy levels as a function of $k/k_c$. The insets show an enlarged region of the $k/k_c$ ranging from 0.9 to 1.1. We set $\Omega=100\omega_0$ and $\xi_\text{R}=0.1\omega_0$. }
 	\label{F2}
 \end{figure}

\begin{figure}[htbp]
	\includegraphics[scale=0.51]{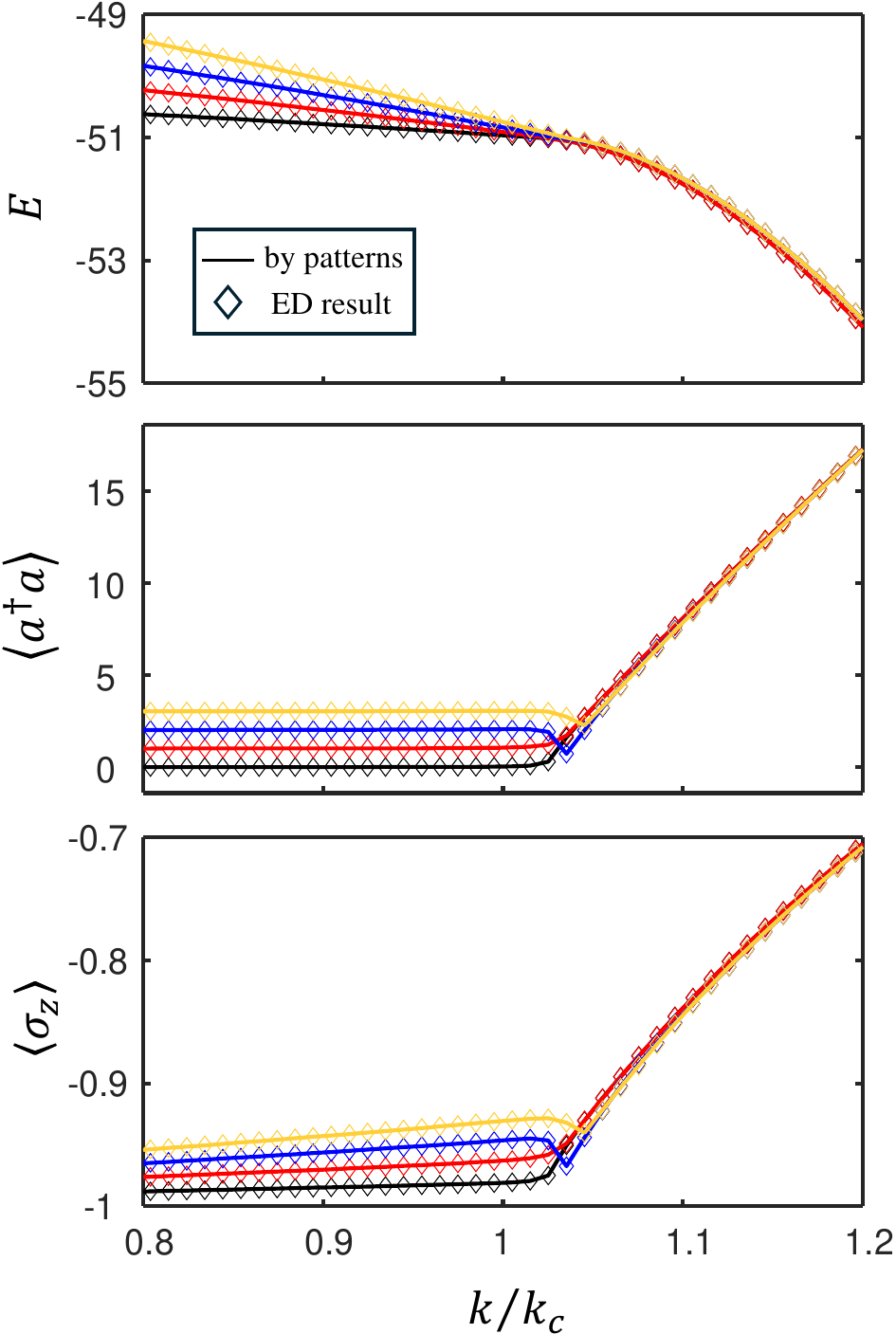}
	\caption{ Comparison between the relevant physical quantities obtained from the patterns and those obtained from ED. (a) Summation of patterns energy levels (lines) and their comparison with the results obtained directly by ED (diamond). (b) Comparison of the summations of the patterns’ photon number (lines) $\langle a^\dagger a\rangle$ to those obtained by ED (diamond). The parameters are the same as Fig.~$\ref{F2}$.}
	\label{F3}
\end{figure}

 We know that a dramatic change in energy at the critical point signifies the occurrence of a QPT, which is a well-known result in Ref.~\cite{PhysRevLett.115.180404,Wang_2022}. We set $\xi_{\text{CR}}=k\xi_{\text{R}}$, i.e., the coupling strength of the counter-rotating term is $k$ times that of the rotating-wave term. Figs~$\ref{F2}$(a)--~$\ref{F2}(c)$ present the first four energy levels against the $k/k_c$ of the three patterns, where $k_c$ is the value of $k$ when $\xi_c=1$ in Eq.~$(\ref{eq6})$.
The energy of all three patterns undergoes a sudden change at the critical point $k/k_c\sim1$. To verify the validity of the patterns, we compare the results of the sum of the corresponding physical quantities energy $E$ and photon number $\langle a^\dagger a\rangle$ of patterns with those obtained directly by numerical exact diagonalization (ED). Figures~$\ref{F3}$(a) and~$\ref{F3}$(b) show sharp changes in both energy levels and photon number near $k/k_c\sim1$, consistent with the results of the superradiant phase transition demonstrated in Ref.~\cite{PhysRevA.95.013819}. Moreover, the results of the summations of the patterns exhibit an exact agreement with those obtained through ED.  

In Fig.~$\ref{F3}$, solid lines denote the summed contributions of the patterns, while symbols indicate the ED results. This agreement stems from the absence of additional approximations in the pattern pictrue compared to the ED method. Consequently, this formalism enables a systematic decomposition of the anisotropic quantum Rabi model Hamiltonian into fundamental patterns. Such a decomposition provides a new perspective  for analyzing phenomena in the anisotropic quantum Rabi model, such as the superradiant phase transition central to this manuscript.
\subsection{The properties of phase transition}
The anisotropic quantum Rabi model exhibits a superradiant phase transition at the critical ratio $k/k_c\sim 1$, where the energy levels, mean photon number  $\langle a^\dagger a\rangle$ and the average value of the spin component $\langle\sigma_z\rangle$ undergo abrupt changes, as shown in Fig.~$\ref{F3}$.

 These abrupt changes are characteristic features of the system's transition from the normal phase to the superradiant phase, which is consistent with the universal critical behavior of light-matter systems near the quantum critical point~\cite{PhysRevLett.115.180404}. We systematically analyze the origin of the phase transition by quantifying the contributions of distinct patterns to the ground and first excited states. This analysis reveals a competition mechanism among the patterns, and we find that the ratio $k$ influences the competition between patterns.\begin{flushleft}
	
\end{flushleft} 

\begin{figure}
	\includegraphics[scale=0.382]{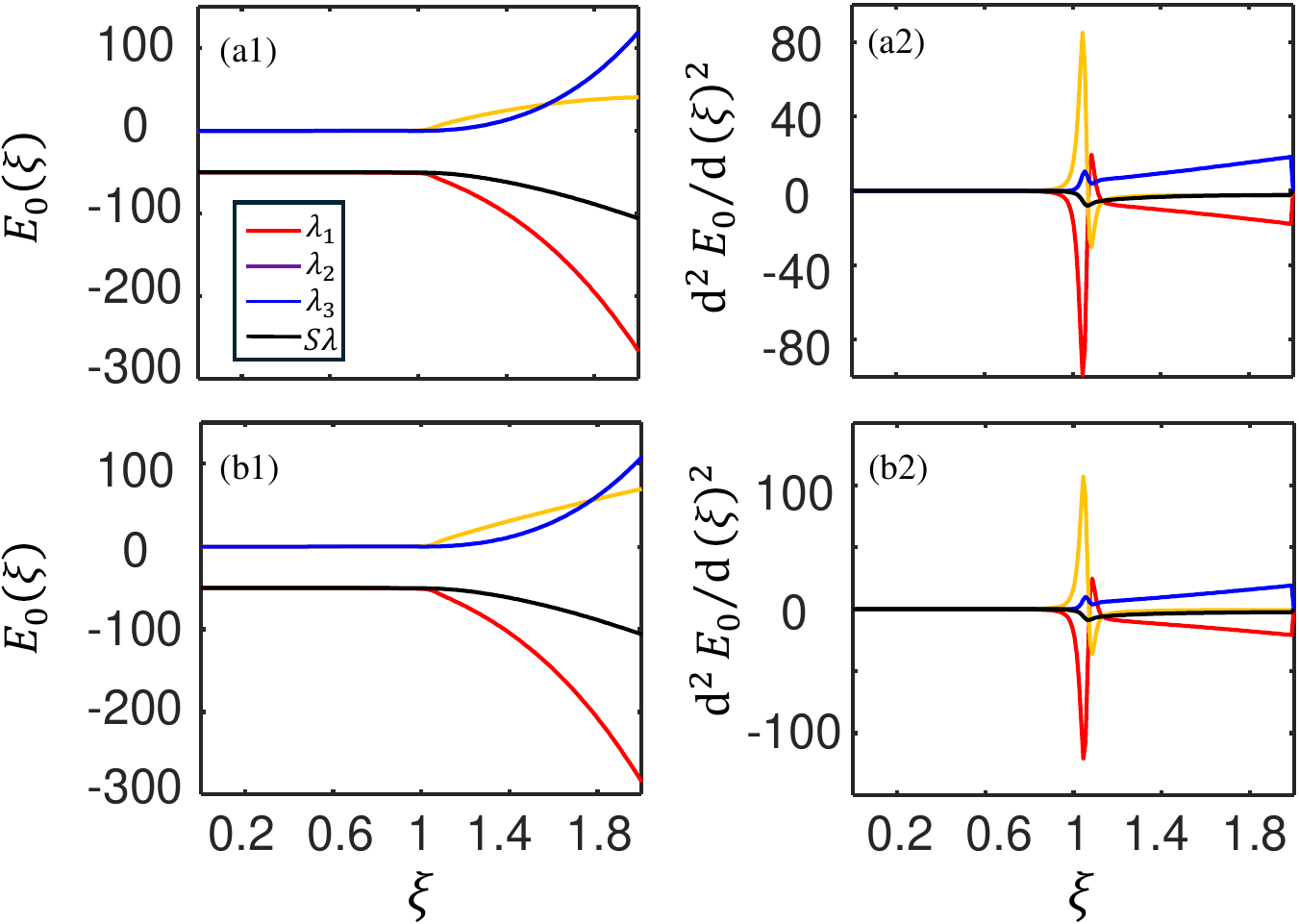}
	\caption{(a1),(b1)  Sum of patterns of ground state energy level (black solid line) and corresponding pattern components (red, yellow, and blue solid lines) as functions of coupling strength $\xi$ ($\xi=\xi_{\text{R}}/\xi^c_{\text{R}}$) when $k=0.9$ and $k=1$. (a2),(b2) Second-order
		derivatives of the corresponding energy levels (black solid line) and corresponding pattern components (red, yellow, and blue solid lines).}
	\label{F4}
\end{figure}

The upper panel of Fig.~$\ref{F4}$ corresponds to the contribution of patterns to the energy level of ground state and its second derivative as a function of $\xi$ ($\xi=\xi_{\text{R}}/\xi^c_{\text{R}}$, where $\xi^c_{\text{R}}$ is the value of $\xi_{\text{R}}$ when $\xi_c=1$ in Eq.~$(\ref{eq6})$) when $k=0.9$. As shown in Fig.~$\ref{F4}$(a1), in the weak-coupling regime ($\xi\leq 1$), the ground state is entirely governed by the dominant pattern $\lambda_1$ (red solid line), while contributions from other patterns ($\lambda_2$, $\lambda_3$) are negligible. This regime corresponds to the well-known normal phase of the anisotropic quantum Rabi model, characterized by a zero photon number.  As the coupling strength increases, the system enters the superradiant phase, where the different patterns exhibit the following distinct critical behaviors. 

(i) Descend of $\lambda_1$. The pattern $\lambda_1$ experiences a sharp decline followed by a continuous descend. (ii) Compensatory response of $\lambda_2$. The pattern $\lambda_2$ exhibits a rapid compensatory growth in response to the decline of $\lambda_1$ (iii) Response to the decline of $\lambda_3$. The imbalance between 
$\lambda_1$ and $\lambda_2$ triggers the $\lambda_3$, which becomes the primary contributor to stabilize the superradiant phase. This behavior is identical to the competition observed in the lower panels for $k=1$, corresponding to the isotropic Rabi model. In Fig.~$\ref{F4}$(a2) and Fig.~$\ref{F4}$(b2), the second-order derivatives of the contribution of the pattern $\lambda_1$ and the pattern $\lambda_3$ show similar changes but opposite trends as the coupling strength $\xi$ increases in the superradiant phase, highlighting the competition between these patterns. The competition between the patterns $\lambda_1$ and $\lambda_{2,3}$ elucidates the microscopic mechanism of the superradiant phase transition, i.e., by increasing the coupling strength, the photon number is excited, thereby reducing the energy of the system driven by the pattern $\lambda_1$. To counterbalance this energy reduction, the pattern $\lambda_2$ and pattern $\lambda_3$ generate positive energy contributions, stabilizing the system in the emergent superradiant phase. This is related to the characteristics of the patterns (see the appendix for details).

\begin{figure}
	\includegraphics[scale=0.37]{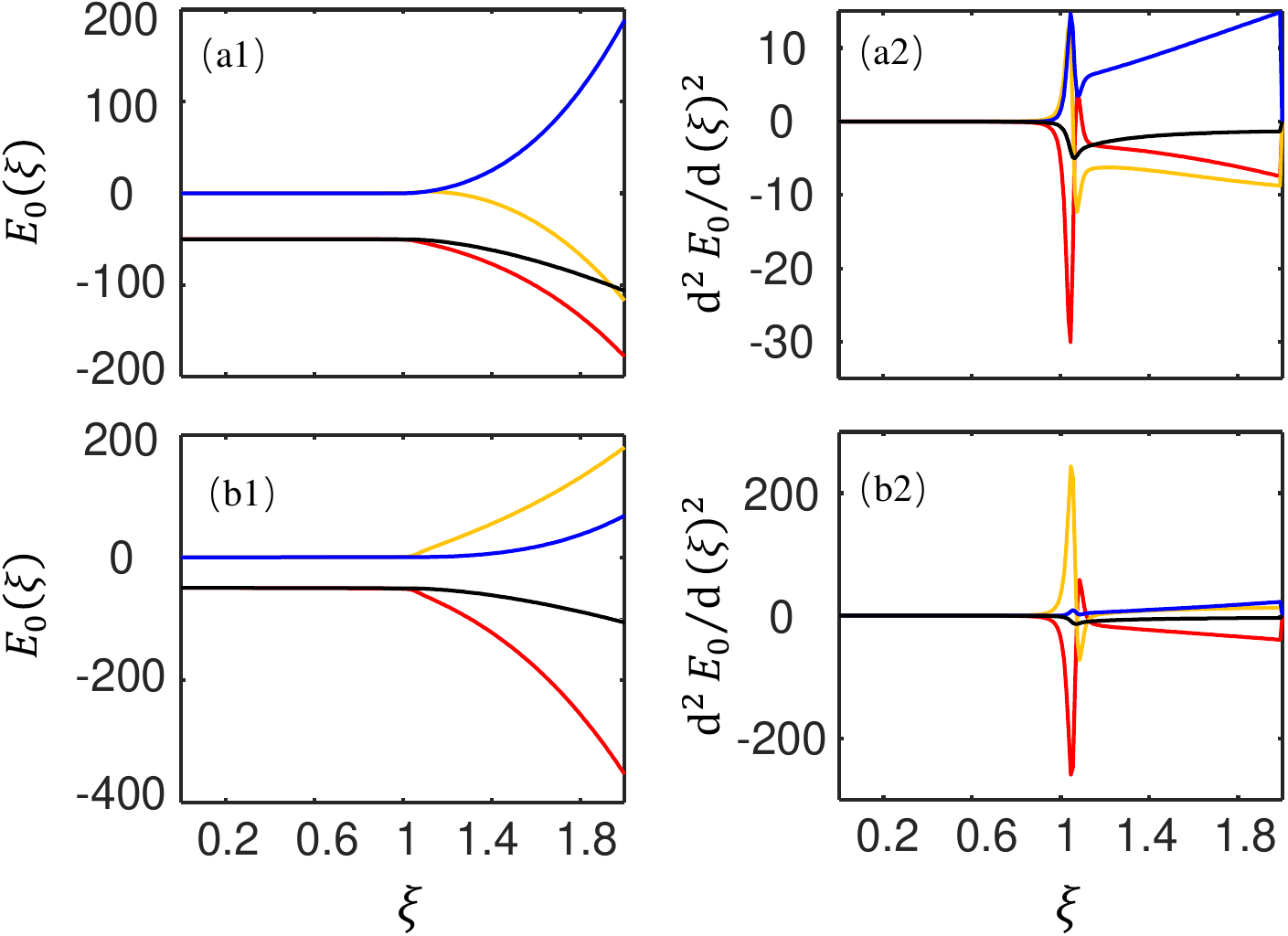}
	\caption{(a1), (b1)  Sum of patterns of ground state energy level (black solid line) and corresponding pattern components (red, yellow, and blue solid lines) as functions of coupling strength $\xi$ ($\xi=\xi_{\text{R}}/\xi^c_{\text{R}}$) when $k=0.5$ and $k=1.5$. (a2), (b2) Second-order
		derivatives of the corresponding energy levels (black solid line) and corresponding pattern components (red, yellow, and blue solid lines).}
	\label{F5}
\end{figure}

\begin{figure}[htbp]
	\includegraphics[scale=0.423]{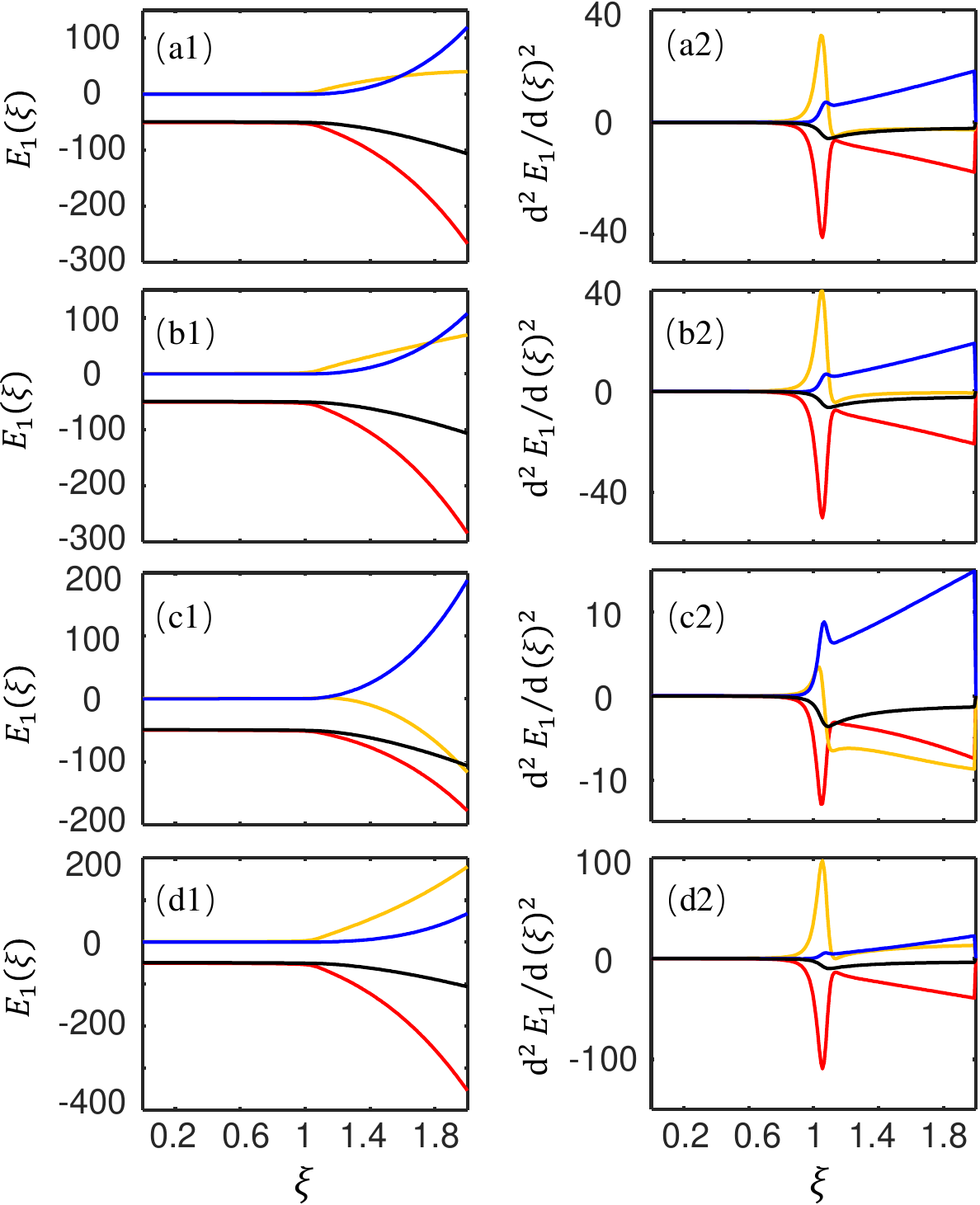}
	\caption{(a1)-(d1)  Sum of patterns of first excited state energy level (black solid line) and corresponding pattern components (red, yellow, and blue solid lines) as functions of coupling strength $\xi$ ($\xi=\xi_{\text{R}}/\xi^c_{\text{R}}$) when $k=0.9$, 1, 0.5 and 1.5. (a2)-(d2) Second-order
		derivatives of the corresponding energy levels (black solid line) and corresponding pattern components (red, yellow, and blue solid lines).}
	\label{F6}
\end{figure}

However, for $k=0.5$ and 1.5, the competition between patterns exhibits different behavior in the superradiant phase. When $k=0.5$, both the patterns $\lambda_1$ and $\lambda_2$ decrease, with growing coupling strength. At the same time, the pattern $\lambda_3$ develops a compensatory response that balances their energy reduction [see Fig.~$\ref{F5}$(a1)]. When $k=1.5$, as the coupling strength increases, the pattern $\lambda_1$ exhibits continuous suppression. The pattern $\lambda_2$ provides a compensatory response but cannot fully balance the energy reduction caused by the pattern $\lambda_1$, making the pattern $\lambda_3$ becomes crucial for balancing the pattern $\lambda_1$, ultimately compensating for its energy loss [see Fig.~$\ref{F5}$(b1)]. Unlike what is shown in Fig.~$\ref{F4}$, the pattern $\lambda_2$ here compensates more energy than the pattern $\lambda_1$. The interplay of energy suppression ($\lambda_1$) and compensation ($\lambda_{2,3}$) directly manifests the critical competition governing the phase transition, providing a unified framework for understanding the emergence of superradiant phase in anisotropic QRM. Similar phenomena can be observed in the first excited state, as shown in Fig.~$\ref{F6}$.

\begin{figure}
	\includegraphics[scale=0.28]{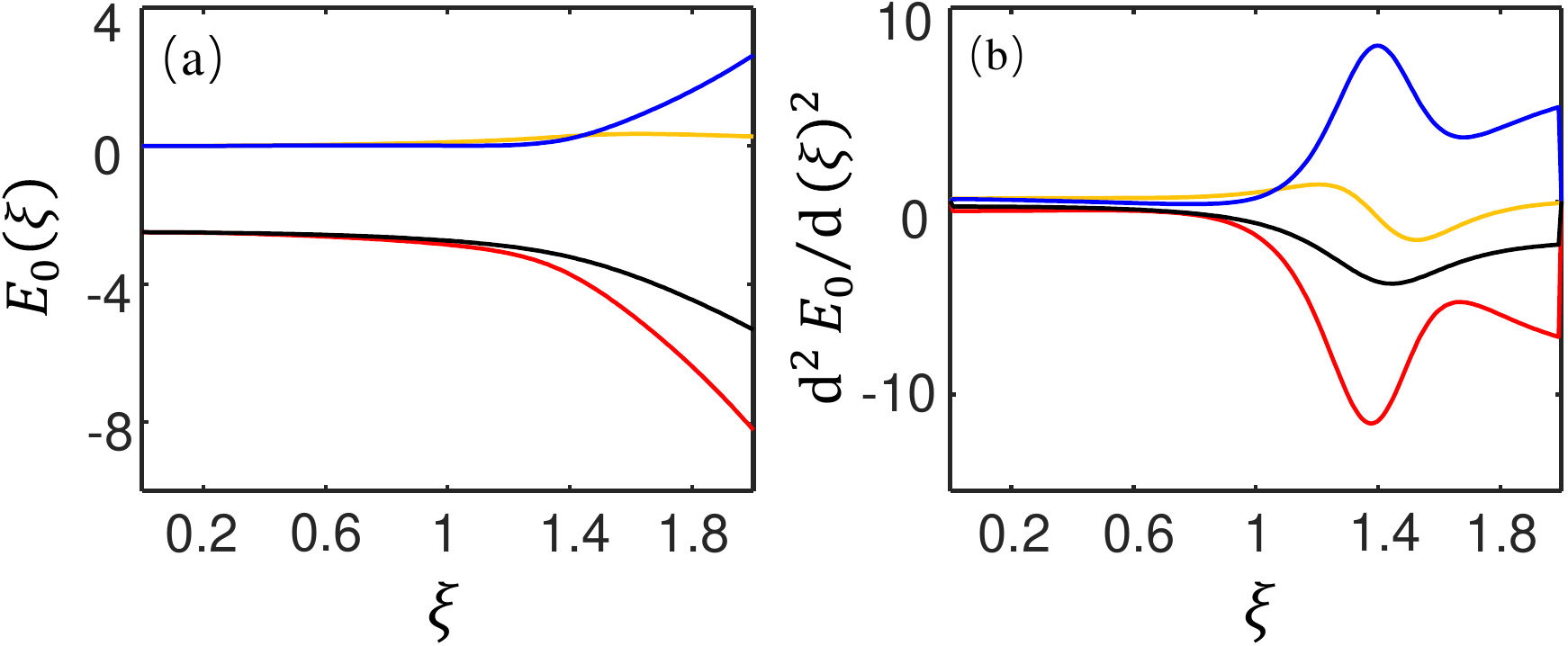}
	\caption{(a) Sum of patterns of ground state energy
		level (black solid line) and corresponding pattern components
		(red, yellow, and blue solid lines) as functions of coupling strength $\xi$ ($\xi=\xi_{\text{R}}/\xi^c_{\text{R}}$) in the regime where the classical-oscillator limit is not satisfied. (b) Second-order derivatives of the corresponding energy levels
		and corresponding pattern components as functions of coupling strength $\xi$ in the regime where the classical-oscillator limit is not satisfied. We set $\Omega=5\omega_0$ and $\xi_{\text{CR}}=0.9\xi_{\text{R}}$.}
	\label{F7}
\end{figure}

 We then explore how the pattern competition evolves when the classical oscillator limit is not satisfied. Taking $k=0.9$ as an example, we simulate the contributions of each pattern to the ground-state energy and their second derivatives with respect to the coupling strength $\xi$. In Fig.~$\ref{F7}$(a), we observe that the pattern $\lambda_1$ begins to decline before the critical coupling. Meanwhile, the pattern $\lambda_2$ also initiates a compensatory response before the critical point, but it is insufficient to fully counter the energy loss from the pattern $\lambda_1$. Ultimately, the pattern $\lambda_3$ is responsible for compensating for the energy reduction induced by the pattern $\lambda_1$. This is because, in a system that does not satisfy the classical oscillator limit, the Hilbert space is effectively a finite-dimensional, truncated space. As the coupling strength increases, the system intends to lower its energy by exciting photons. However, this process is constrained by the truncated space. Therefore, pattern $\lambda_1$ will respond in advance, causing the system's energy to decrease. The same characteristic is also observed in the second derivative of the ground-state energy [see Fig.~$\ref{F7}$(b)].

\subsection{Patterns and fidelity out-of-time-order correlations}

 To further demonstrate the value of the pattern decomposition, we analyze the behavioral associations between these patterns and out-of-time-order correlation function (OTOC). 
	
	The out-of-time-order correlation functions (OTOCs) are defined as~\cite{Sun2020,PhysRevA.105.032444}                  
	\begin{eqnarray}
		\mathcal{F}(t) =\langle \hat{W}^{\dagger}(t)\hat{V}^\dagger \hat{W}(t)\hat{V}\rangle.
	\end{eqnarray}
	Here, the angular brackets denote averaging over the initial state $|\psi_0\rangle$, $\hat{W}$ and $\hat{V}$ are two commuting Hermitian operators at initial time $(t=0)$, i.e., $[\hat{W},\hat{V}]=0$, where $\hat{W}(t) = e^{i\hat{H}_{AN}t} \hat{W}e^{-i\hat{H}_{AN}t}$. We consider $\mathcal{F}(t)$ as a fidelity OTOC (FOTOC) under the condition that the initial state $|\psi_0\rangle$ is an eigenstate of $\hat{V}$ and $\hat{W}_G=e^{i\delta_{\phi}\hat{G}}$ with $\hat{G}$ being Hermitian and $\delta_{\phi}$ a small perturbation parameter. We define $\hat{V}$ as the projector onto the initial state, $\hat{V} = \hat{\rho}(0) = |\psi_0\rangle\langle \psi_0|$, with $|\psi_0\rangle = |+,0\rangle$. Given that $\delta_{\phi}$ represents a small perturbation, the fidelity OTOC $\mathcal{F}_G(t) =\langle \hat{W}^{\dagger}(t)\hat{\rho}(0)\hat{W}(t)\hat{\rho}(0)\rangle$ can be expanded as a power series in $\delta_{\phi}$, leading to 
	\begin{eqnarray}
		1-\mathcal{F}_G(t) =\delta_{\phi}^2(\langle\hat{G}^2(t) \rangle - \langle\hat{G}(t)\rangle^2) = \delta_{\phi}^2 \text{var} \hat{G}(t).
	\end{eqnarray}
	
\begin{figure}
	\includegraphics[scale=0.376]{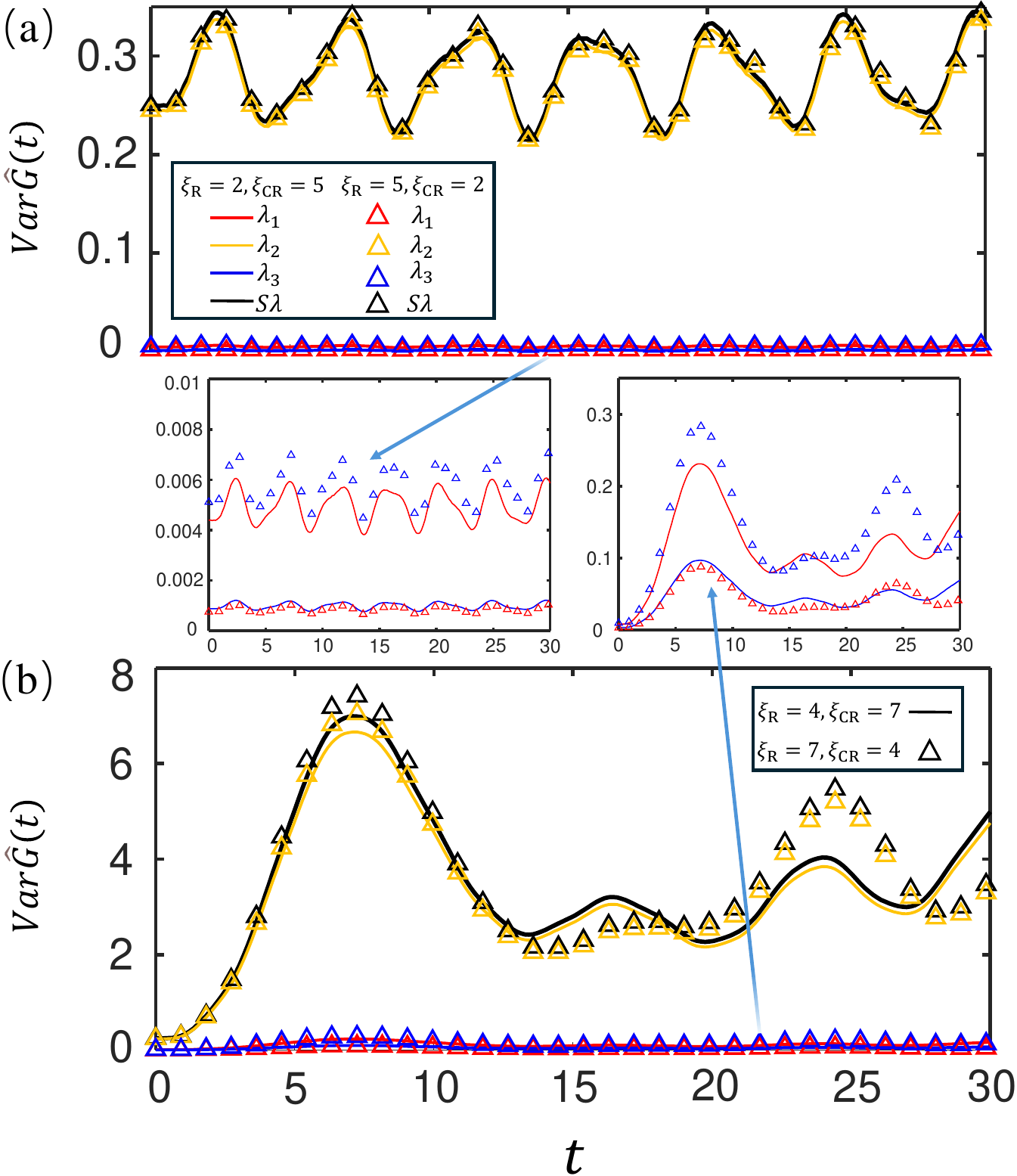}
	\caption{Evolution of FOTOC for the anisotropic Rabi model with Hamiltonian (1). (a) Evolution of FOTOC in the normal phase and its pattern contributions for the cases of $\xi_{\text{R}}=2, \xi_{\text{CR}}=5$ and $\xi_{\text{R}}=5, \xi_{\text{CR}}=2$. (b) Evolution of FOTOC in the superradiant phase and its pattern contributions for the cases of $\xi_{\text{R}}=4, \xi_{\text{CR}}=7$ and $\xi_{\text{R}}=7, \xi_{\text{CR}}=4$. We set $\Omega=100\omega_0$, i.e., the critical value of $\xi_c=\xi_{\text{R}}+\xi_{\text{CR}}=10$.}
	\label{F8}
\end{figure}

	As shown in Fig.~$\ref{F8}$, we simulate the variance of $\hat{G}=(\hat{a}^{\dagger}+\hat{a})/2$. The simulation results indicate that in the normal phase $(\xi_{\text{R}}+\xi_{\text{CR}}<10)$\,[see Fig.~$\ref{F8}$(a)], the oscillation amplitude of the FOTOC remains almost constant, whereas in the superradiant phase $(\xi_{\text{R}}+\xi_{\text{CR}}>10)$\,[see Fig.~$\ref{F8}$(b)], it exhibits rapid growth. Furthermore, a key finding is that the FOTOC is overwhelmingly dominated by the pattern $\lambda_2$ in both phases (the yellow solid line and the triangles in Fig.~$\ref{F8}$), with the patterns $\lambda_1$ and $\lambda_3$ making only minimal contributions. This is because the fluctuations of the photonic quadrature are dominated by the pattern $\lambda_2$ (see the appendix for details). Fidelity Out-of-Time-Order Correlator (i.e., FOTOC), it is a quantum information metric used to diagnose quantum chaos, information scrambling, and phase transitions~\cite{Lewis-Swan2019}. The early-time behavior of the FOTOC directly reflects the strength of quantum fluctuations of the ground-state wave function along a specified operator direction, such as $(a+a^{\dagger})$~\cite{Kirkova2023}. The expectation value of the operator $(a+a^{\dagger})$ serves as an order parameter in the superradiant phase transition (changing from zero to nonzero), while its fluctuations (variance) diverge at the critical point, serving as a key observable characterizing the critical behavior of the phase transition. The contribution from the pattern $\lambda_1$ to the FOTOC exceeds that of the pattern $\lambda_3$ when $\xi_{\text{R}}<\xi_{\text{CR}}$, while this trend is reversed for $\xi_{\text{R}}>\xi_{\text{CR}}$. This analysis offers compelling support for the interpretation that the underlying mechanism of the superradiant phase transition in the anisotropic Rabi model is the competition and collaboration between the patterns.

The classical oscillator limit ($\Omega/\omega_0 \to \infty$), crucial for studying quantum phase transitions in the anisotropic Rabi model, is experimentally challenging to realize. To address this, we propose a parametric amplification method in a driven cavity QED system. This method exponentially enhances the coupling strength while suppressing the effective cavity frequency, thereby effectively realizing the classical oscillator limit and enabling access to anisotropic Rabi model phase transitions under experimentally feasible conditions.

\section{Model and Hamiltonian}\label{s4}

\begin{figure}
	\includegraphics[scale=0.63]{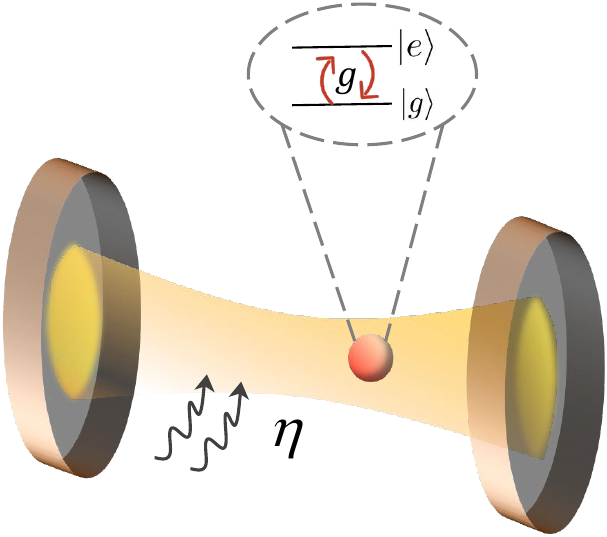}
	\caption{Schematic of a qubit weakly coupled to a cavity driven by two-photon, with an amplitude $\eta$. The states $\left|g\right>$ and $\left|e\right>$ are the ground and excited states of the qubit, respectively. The qubit-cavity coupling strength is $g$.}
	\label{F9}
\end{figure}

The system of a qubit weakly coupled to a cavity driven by the two-photon drive can be illustrated as Fig.~$\ref{F9}$. 
\begin{figure}
	\includegraphics[scale=0.56]{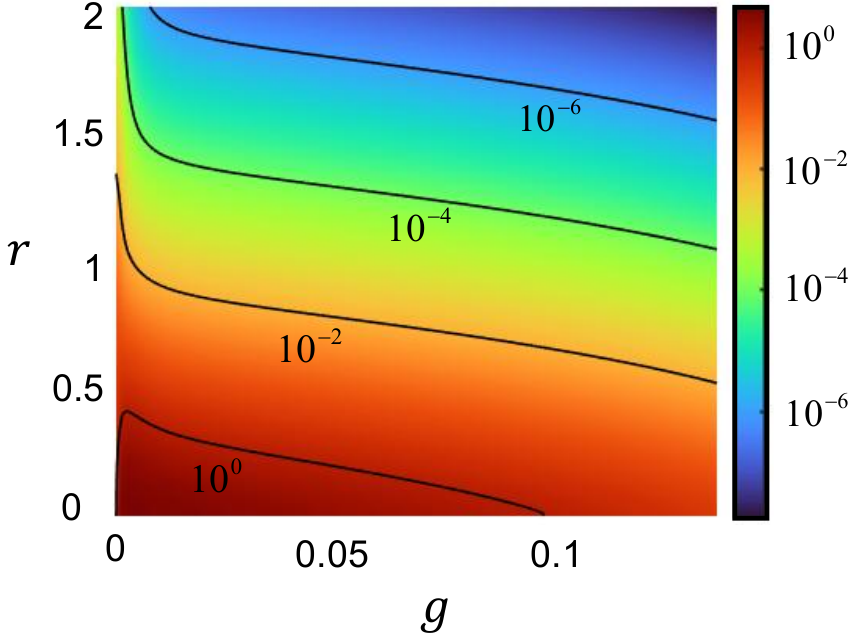}
	\caption{Thermogram shows the distribution of the logarithmic value $\mathrm{log}(E_1-E_0)$ of the energy gap between the excited state and the ground state in the parameter space $(g,r)$.}
	\label{F10}
\end{figure}
\begin{figure}
	\includegraphics[scale=0.53]{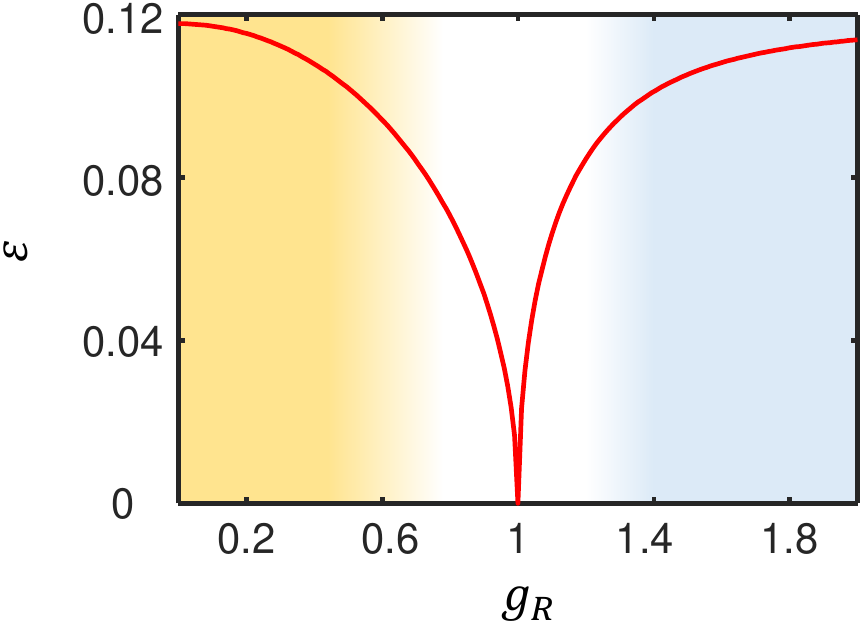}
	\caption{Excitation energies of the anisotropic Rabi Hamiltonian as functions of the coupling strengths $g_R$ ($g_R=g/g_0$). We choose parameters $r=\sqrt{2}$ and we set $\delta_q=200\delta_c\mathrm{sech}(2r)=23.56\delta_c$. The disappearance of $\epsilon$ at the critical point $g_R$ signifies the emergence of the QPT. Here, $g_0=\sqrt{\delta_c\mathrm{sech}(2r)\delta_q}/e^r$.}
	\label{F11}
\end{figure}
\begin{figure}
	\includegraphics[scale=0.5]{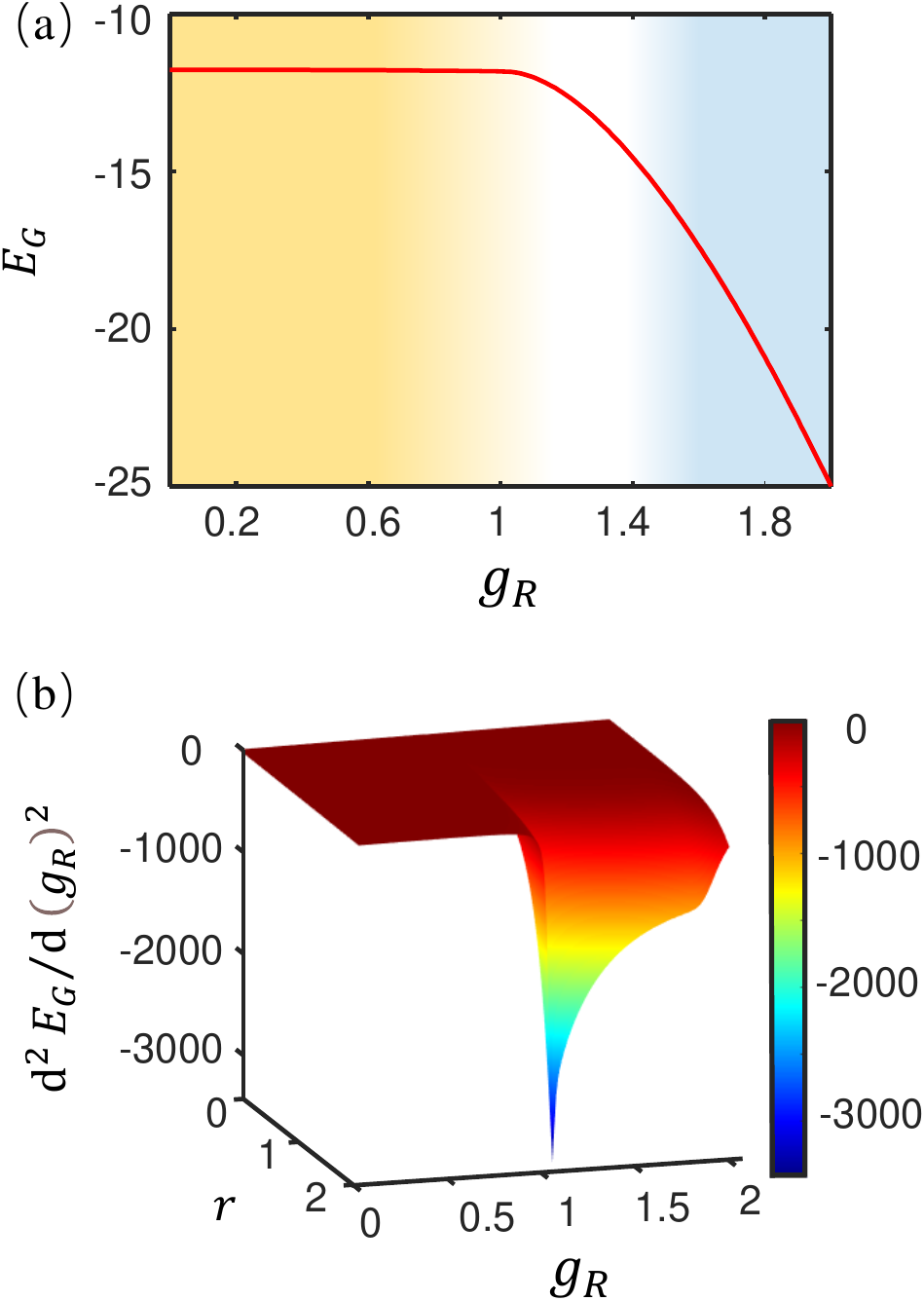}
	\caption{ (a) Ground-state energy $E_{G}$ as the function of the coupling strengths $g_R$ ($g_R=g/g_0$). (b) The second
		derivative of ground-state energy $d^2E_{G}/d(g_R)^2$ as the function of the coupling strengths $g_R$. We choose parameters $r=\sqrt{2}$ and we set $\delta_q=200\delta_c\mathrm{sech}(2r)=23.56\delta_c$. Here, $g_0=\sqrt{\delta_c\mathrm{sech}(2r)\delta_q}/e^r$.}
	\label{F12}
\end{figure}
\begin{figure}
	\includegraphics[scale=0.56]{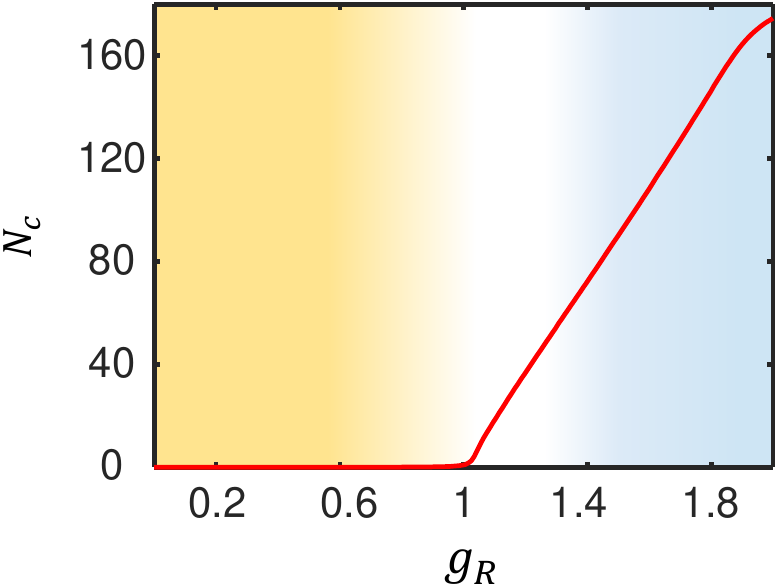}
	\caption{Photon number of the anisotropic Rabi Hamiltonian as functions of the coupling strengths $g_R$ ($g_R=g/g_0$). We choose parameters $r=\sqrt{2}$ and we set $\delta_q=200\delta_c\mathrm{sech}(2r)=23.56\delta_c$. The nonzero $N_c$ at the critical point $g_R=1$ indicates the occurance of the QPT. Here, $g_0=\sqrt{\delta_c\mathrm{sech}(2r)\delta_q}/e^r$.}
	\label{F13}
\end{figure}
Working in a frame rotating at half the parametric drive frequency $\omega_p/2$, the Hamiltonian can be written as
\begin{eqnarray}
	\begin{aligned}
		H=&\delta_c a^{\dagger}a+\frac{\delta_q}{2}\sigma_z-\frac{\eta}{2}(a^{\dagger 2}+a^2)\\
		&+g(a^{\dagger}\sigma_- + a\sigma_+),\label{eq22}
	\end{aligned}
\end{eqnarray}
where $\sigma_\pm=(\sigma_x\pm i\sigma_y)/2$ are the atomic transition operators with the Pauli matrices $\sigma_{x,y,z}$. While $a$ $(a^{\dagger})$ is the annihilation (creation) operator of the cavity, $\eta$ is the parametric drive amplitude and $g$ is the qubit-cavity coupling strength. The $\delta_{c}=\omega_{c}-\omega_p/2$ is the cavity detuning, and $\delta_{q}=\omega_{q}-\omega_p/2$ is the qubit detuning (with $\omega_{c}$ and $\omega_{q}$ being the cavity and qubit frequencies, respectively).

The Hamiltonian $H$ in Eq.~$(\ref{eq22})$ can be diagonalized by the squeezed operator $S(r)=\exp[r(a^2-a^{\dagger 2})/2]$, where $r$ is the squeezing parameter, defined by $\tanh2r=\eta/\delta_c$. The squeezing transformation leads to 
\begin{eqnarray}
	\begin{aligned}
		H_{\mathrm{AR}}&\equiv S(r)HS^{\dagger}(r)=H_0+H_1,
	\end{aligned}\label{eq2} 
\end{eqnarray}
where
\begin{eqnarray}
	\begin{aligned}
		H_0&=\delta_c \mathrm{sech}(2r)a^{\dagger}a+\frac{\delta_q}{2}\sigma_z,\\
		H_1&=g_1(a\sigma_++a^{\dagger}\sigma_-)+g_2(a\sigma_-+a^{\dagger}\sigma_+).
	\end{aligned}
\end{eqnarray}
Here, $g_1=g \mathrm{cosh}2r$ and $g_2=g \mathrm{sinh}2r$  denote the qubit-cavity coupling strengths of the rotating-wave and counter-rotating-wave interactions, respectively.
The Hamiltonian $H_{\mathrm{AR}}$ in Eq.~$(\ref{eq2})$ is a simulated anisotropic Rabi Hamiltonian.  It is distinctly different from the isotropic Rabi model with $g_1=g_2$ which has been proven to exhibit the QPT phenomenon~\cite{PhysRevLett.115.180404,chen2024sudden}.

We denote $|\uparrow (\downarrow)\rangle$ as eigenstates of the qubit, and Fock states $|m\rangle$ ($m=0,1,2,3,...$) as the states of the cavity. The Hamiltonian $H_0$ has decoupled spin subspaces $\Gamma_{(\downarrow)}$ and $\Gamma_{(\uparrow)}$. For $\delta_c\mathrm{sech}(2r)/\delta_q\rightarrow \infty$, the eigenstates of the lowest energy of $H_0$ correspond to the eigenstates of the resonator confined in $\Gamma_{(\downarrow)}$. However, the interaction Hamiltonian $H_1$ introduces the coupling between the two spin subspaces, which makes the virtual excitation between them alter the nature of the low-energy eigenstates and eigenenergies. Therefore, we employ the time-dependent perturbation method~\cite{PhysRevA.95.032124} to obtain the effective Hamiltonian, which has no coupling between $\Gamma_{(\downarrow)}$ and $\Gamma_{(\uparrow)}$.

In a frame rotating with $\frac{\delta_q}{2}\sigma_z$, the Hamiltonian $H_{\mathrm{AR}}$ becomes
\begin{eqnarray}
	\begin{aligned}
		H^{\prime}=&\delta_c \mathrm{sech}(2r)a^{\dagger}a+g_1(a\sigma_+e^{i\delta_q t}+a^{\dagger}\sigma_- e^{-i\delta_q t})\\&+g_2(a\sigma_-e^{-i\delta_q t}+a^{\dagger}\sigma_+ e^{i\delta_q t})+\frac{\delta_q}{2}\sigma_z ,
	\end{aligned}\label{eq4} 
\end{eqnarray}
then, according to the time-dependent perturbation method~\cite{PhysRevA.95.032124}, the Hamiltonian $H^{\prime}$ in Eq.~$(\ref{eq4})$ can be written as
\begin{eqnarray}
	\begin{aligned}
		H_I=&\delta_c\mathrm{sech}(2r)a^{\dagger}a+\frac{g_1^2+g_2^2}{\delta_q}a^{\dagger}a\sigma_z+(\frac{g_1^2}{\delta_q}+\frac{\delta_q}{2})\sigma_z\\&+\frac{g_1g_2}{\delta_q}(a^{\dagger 2}+a^2)\sigma_z+\frac{g_1^2-g_2^2}{\delta_q}\sigma_-\sigma_+.
	\end{aligned}
\end{eqnarray}

We use the same method as in Ref.~\cite{PhysRevA.95.013819} to derive the following quantities for both normal and superradiant phases: low-energy Hamiltonians, excitation energies, eigenvalues with corresponding eigenstates, and critical coupling strength. The low-energy effective Hamiltonian of the normal phase takes the form
\begin{eqnarray}
	\begin{aligned}
		H_{\mathrm{np}}=&(\delta_c\mathrm{sech}(2r)-\frac{g_1^2+g_2^2}{\delta_q})a^{\dagger}a\\&-\frac{g_1g_2}{\delta_q}(a^{\dagger 2}+a^2)-\frac{g_1^2}{\delta_q}-\frac{\delta_q}{2}.
	\end{aligned}
\end{eqnarray}
The squeezing transformation $S'^{\dagger}(r'_{\mathrm{np}})H_{\mathrm{np}}S'(r'_{\mathrm{np}})$ yielding $H'_{\mathrm{np}}=\epsilon_{\mathrm{np}}a^{\dagger}a+E_{\mathrm{np}}$, with the excitation energy
\begin{eqnarray}
	\epsilon_{\mathrm{np}}
	=\sqrt{\!\Bigl(\delta_c\operatorname{sech}(2r)
		-\tfrac{g^{2}\cosh(2r)}{\delta_q}\Bigr)^{\!2}
		-\tfrac{g^{4}\sinh^{2}(2r)}{\delta_q}},
	\label{eq10}
\end{eqnarray}

and the squeezing parameter
\begin{eqnarray}
	r'_{\mathrm{np}}=\frac{1}{4}\mathrm{ln}[1+\frac{2g^2\mathrm{sinh}(2r)}{\delta_c \mathrm{sech}(2r)\delta_q-g^2e^{2r}}].
\end{eqnarray}
Then we obtain the critical couping strength
\begin{eqnarray}
	g_c=\frac{ge^r}{\sqrt{\delta_c\mathrm{sech}(2r)\delta_q}}\label{eq29},
\end{eqnarray}
the eigenstates and eigenvalues of $H'_{\mathrm{np}}$ are given by
\begin{eqnarray}
	\ket{\phi_{\mathrm{np}}^m} = S'(r'_{\mathrm{np}}) \ket{m} \ket{\downarrow},
\end{eqnarray}
\begin{eqnarray}
	E_{\mathrm{np}}^m=m\epsilon_{\mathrm{np}}+E_{\mathrm{np}}.
\end{eqnarray}

Likewise in the superradiant phase, we obtain the low-energy effective Hamiltonian 
\begin{equation}
	\begin{aligned}
		H_{\mathrm{sp}}\simeq &[\delta_c\mathrm{sech}(2r)-\frac{(g_1+g_2)^2g_c^{-4}+(g_1-g_2)^2}{2\delta_qg_c^2}]a^{\dagger}a\\
		&-\frac{(g_1+g_2)^2g_c^{-4}-(g_1-g_2)^2}{4\delta_qg_c^2}(a^{\dagger 2}+a^2)\\
		&-\frac{[(g_1+g_2)g_c^{-2}-(g_1-g_2)]^2}{4\Omega g_c^2}-\frac{\Omega g_c^{-2}}{2},
	\end{aligned}
\end{equation}
the excitation energy
\begin{equation}
\epsilon_{\mathrm{sp}}
= \sqrt{%
	\!\begin{aligned}[t]
		&\biggl[\delta_{c}\operatorname{sech}(2r)
		-\frac{\delta_{c}\operatorname{sech}(2r)\delta_{q}g_{c}^{-2}
			+g^{2}e^{-2r}}
		{2\delta_{q}g_{c}^{2}}\biggr]^{2} \\[4pt]
		&\quad -\biggl[\frac{\delta_{c}\operatorname{sech}(2r)\delta_{q}
			-g^{2}e^{-2r}}
		{4\delta_{q}g_{c}^{2}}\biggr]^{2}
	\end{aligned}%
},
\end{equation}
and the eigenstates and eigenenergies of $H_{\mathrm{sp}}$
\begin{eqnarray}
	\begin{aligned}
		|\phi_{\text{sp}}^m\rangle = D(\pm \alpha_0) S'(r'_{\text{sp}}) |m\rangle |\downarrow^{\pm}\rangle,
	\end{aligned}
\end{eqnarray}
where
\begin{eqnarray}
	\begin{aligned}
		r'_{\text{sp}} = \frac{1}{4} \ln \left[ \frac{2g^2\mathrm{sinh(2r)}}{g^2e^{2r}} (1 - g_c^{-4})^{-1} \right],	
	\end{aligned}
\end{eqnarray}
\begin{eqnarray}
	\begin{aligned}
		|\downarrow^{\pm}\rangle = \mp \sqrt{\frac{1 - g_c^{-2}}{2}} |\uparrow\rangle + \sqrt{\frac{1 + g_c^{-2}}{2}} |\downarrow\rangle.
	\end{aligned}
\end{eqnarray}

\section{analysis of Quantum PHASE TRANSITION}\label{s5}
 This section provides detailed numerical simulations that demonstrate the effectiveness of our proposed model from Section IV for simulating the QPT in the anisotropic Rabi model. Under the weak coupling mechanism, we can control the occurrence of phase transitions by changing the two-photon driving strength. 
 
 As shown in Fig.~$\ref{F10}$, when the coupling strength is within a certain range, as the squeezing parameter increases, the energy gap between the excited state and the ground state tends to zero, that is, the energy level undergoes degeneracy. Then, we analyze the characteristics of the system in both normal and superradiant phases when the squeezing parameter is set to be $r=\sqrt{2}$. The excitation energies $\epsilon_{\mathrm{np}}$ and $\epsilon_{\mathrm{sp}}$ that can characterize the behavior of the system as a function of coupling strength $g_R$ ($g_R=g/g_{0}$, where $g_0$ is the value of $g$ when $g_c=1$ in Eq.~$(\ref{eq29})$) are shown in Fig.~$\ref{F11}$.
The numerical results indicate that the excitation energies in both normal and superradiant phases vanishes as the coupling strength approaches the critical point. The ground-state energy as a function of coupling strength $g_R$ is displayed in Fig.~$\ref{F12}$(a) and second derivative of ground-state energy $\mathrm{d}^2E_{G}/\mathrm{d}^2(g_R)$ as a function of coupling strength $g_R$ has displayed in Fig.~$\ref{F12}$(b). From Fig.~$\ref{F12}$, we see that the ground-state energy decreases at the critical point $g_R=1$, and it follows a continuous curve. While $\mathrm{d}^2E_{G}/\mathrm{d}^2(g_R)$ displays a discontinuity at the critical point $g_R=1$, which clearly indicates the second-order nature of the QPT. Moreover, this discontinuity becomes more pronounced as the squeezing parameter $r$ increases.  The photon number in the squeezed picture $N_c=\langle a^{\dagger}a\rangle$, which is the order parameter of the QPT, is presented in Fig.~$\ref{F13}$. For $g_R<1$, i.e., in the normal phase, the photon number $n_c$ is zero, while for $g_R>1$, i.e., in the superradiant phase, the photon number $N_c$ becomes infinity. This indicates that when $g_R>1$, the photon number $N_c$ acquires macroscopic occupations.

\section{conclusion}\label{s6}
In conclusion, we obtained three patterns by diagonalizing the anisotropic Rabi model in operator space and simulated the contribution of three patterns to the ground state and excited state. The results show that the three patterns are in competition during the superradiant phase transition. That is, with the increase of coupling strength, (i) the pattern $\lambda_1$ drives the phase transition; (ii) the pattern $\lambda_2$ has a similar response speed to compensate for the pattern $\lambda_1$, but the energy compensation is less than the pattern $\lambda_1$; (iii) the pattern $\lambda_3$ shows a slow response speed, but plays a key role in balancing the pattern $\lambda_1$. This competitive relationship explains why and how the superradiant phase transition occurs. We also explore how different ratios of the rotating-wave to counter-rotating-wave coupling strengths affect the pattern
competition and the contributions of the patterns to the fidelity out-of-time-order correlation function. In addition, we propose a protocol, using a parametrically-driven Jaynes-Cummings model to explore the quantum phase transition. In the squeezed-light frame, this parametrically-driven Jaynes-Cummings model can accurately simulate the dynamics of an ultrastrong-coupling anisotropic Rabi model. The results of numerical simulations demonstrate that this anisotropic Rabi model undergoes a quantum phase transition at the critical point $g_R\sim1$. To be specific, as the coupling strength approaches the critical point, the excitation energy tends to zero; The ground-state energy is continuous at the critical point, while its second derivative to the coupling strength is discontinuous at the critical point; When the coupling strength exceeds the critical point, the photon number increases to infinity. Our work provides a deeper understanding of the microscopic mechanisms behind the superradiant phase transition in the anisotropic quantum Rabi model, and offers a theoretical framework for the realization of controllable quantum phase transitions.

\section*{Acknowledgements}
Y.-H.C. was supported by the National Natural Science Foundation of China under Grant No. 12304390 and 12574386, the National Postdoctoral Overseas Talent Recruitment Program of China, the Fujian 100 Talents Program, and the Fujian Minjiang Scholar Program. Y.X. was supported by the National
Natural Science Foundation of China under Grant No. 62471143, the Key Program of National Natural
Science Foundation of Fujian Province under Grant No.2024J02008, and the project from Fuzhou University
under Grant No. JG2020001-2.   

\section*{Competing interests}
The authors declare that they have no competing interests.  

\appendix
\section*{Appendix:Discussion on the nature of patterns}

The Hamiltonian of the anisotropic Rabi model is decomposed into three patterns 
	\begin{eqnarray}
		H_{\mathrm{AN}}=\sum_{n=1}^{3} \lambda_n A_n^\dagger A_n,
	\end{eqnarray}
	where $A_n=u_{n,x}\sigma_x+u_{n,y}(-i\sigma_y)+u_{n,a}a$ is the pattern operator formed by a linear combination of $\{\sigma_x,-i\sigma_y,a\}$. These patterns are essentially a linear combination of a set of single-particle operators, and the corresponding eigenvalues $\lambda_n$ are the “weights” of these operator combinations in the Hamiltonian. Their physical meaning is that $\lambda_n$ serves as the “effective energy coefficient” of each pattern, indicating the extent to which this pattern contributes to the total energy of the system. The $\lambda_1$ and $\lambda_3$ mainly contribute to the energy of the atomic degrees of freedom, while the $\lambda_2$ mainly contributes to the energy of the photonic field. The $u_{n,j}$ determines the weights and relative phases of the two distinct atomic degrees of freedom ($\sigma_x,i\sigma_y$) and the photonic degree of freedom ($a$) within each pattern (the $u_{n,x}$, $u_{n,y}$, and $u_{n,a}$ represent the weights of $\sigma_x$, $-i\sigma_y$, and $a$ in the pattern operator $A_n$, respectively). The magnitude of $u_{n,j}$ reflects whether the pattern is mainly dominated by the photon pattern or the atomic pattern. The sign (positive or negative) of $u_{n,j}$ determines the relative phase between different components within the pattern.
	
\begin{figure}
		\includegraphics[scale=0.485]{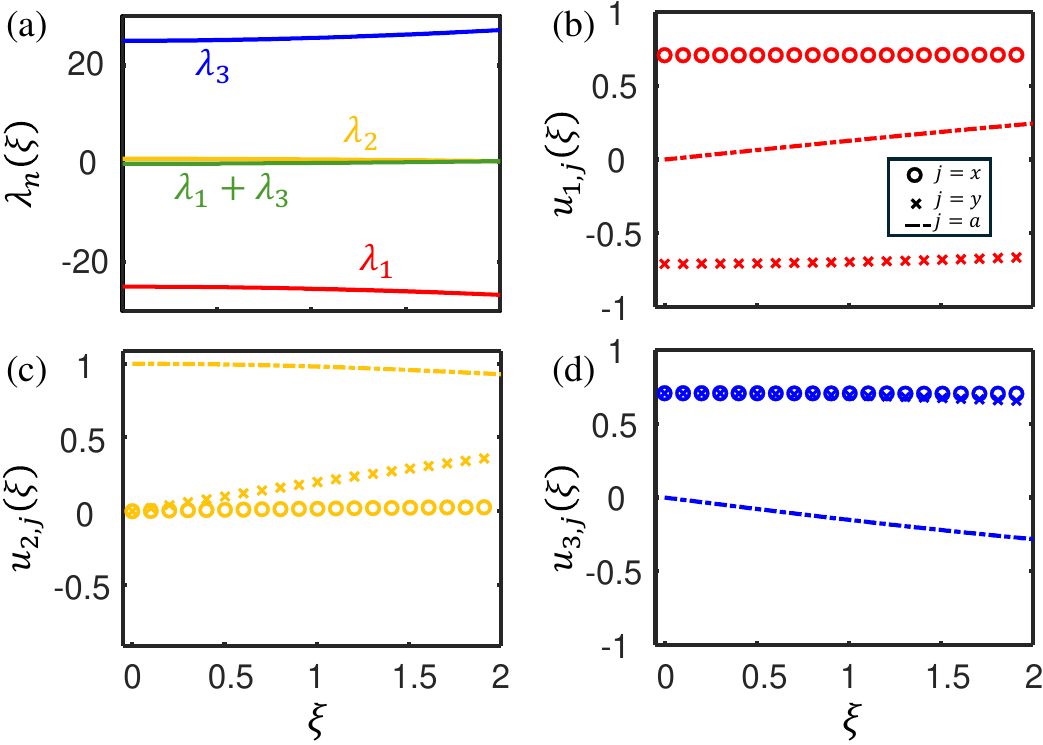}
		\caption{ Patterns characterized by their eigenenergies and eigenvectors as functions of coupling strength $\xi$. (a) the eigenenergy of the pattern $\lambda_n\,(n=1,2,3)$. (b)-(d) eigenvectors $u_{n,j}$\,($n=1,2,3$ and $j=x,y,a$). We set $k=0.9$, the other parameters are the same as Fig.~$\ref{F2}$.}
		\label{F14}
	\end{figure}

	In Fig.~$\ref{F14}$, taking $k=0.9$ as an example, i.e., $\xi_{\text{CR}}=0.9\xi_{\text{R}}$, we simulate the behaviors of $\lambda_n\,(n=1,2,3)$ and $u_{n,j}\, (j=x,y,a)$ with respect to the anisotropy parameter $\xi\,(\xi=\xi_{\text{R}}/\xi^c_{\text{R}})$. From Fig.~$\ref{F14}$(b), in pattern $\lambda_1$, the weights of the spin operator components $\sigma_x$ and $-i\sigma_y$ (i.e., $u_{1,x}$, $u_{1,y}$) are significantly larger than that of the photonic component $a$ ($u_{1,a}$), indicating that $\lambda_1$ is primarily “spin patterns”. As the coupling strength increases, the weight of the photon component increases. This enhances the interaction between the photon pattern and the spin pattern, thereby redistributing the ground-state energy of the system and accounting for the energy reduction driven by $\lambda_1$.
	\begin{figure}
		\includegraphics[scale=0.73]{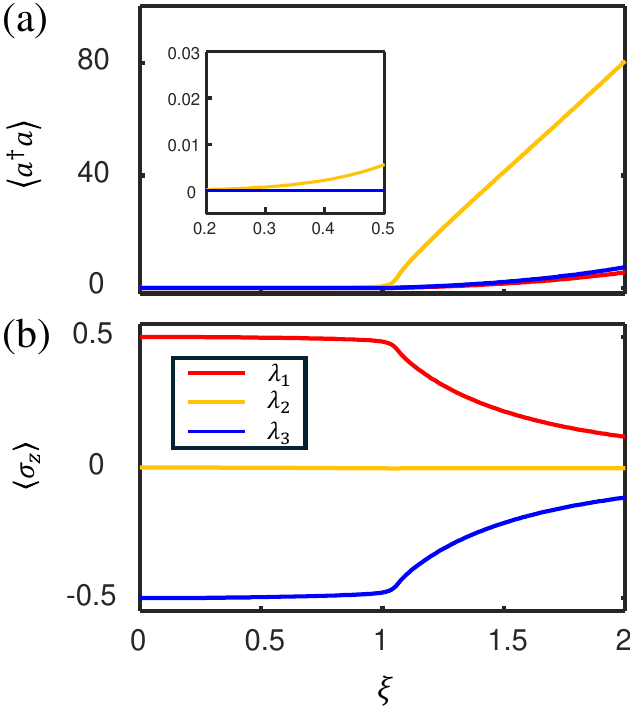}
		\caption{ Photon number (a) and the spin components (b) of the ground state in the corresponding patterns as functions of the coupling strength $\xi$ when $k=0.9$. The other parameters are the same as Fig.~$\ref{F2}$.}
		\label{F15}
	\end{figure}
	
	As shown in Fig.~$\ref{F14}$(c), the pattern $\lambda_2$ is dominated by the photon component $a$ (i.e., $u_{2,a}$), thus it represents the photonic pattern. Since the pattern $\lambda_2$ encompasses most of the photonic degrees of freedom, when the pattern $\lambda_1$ drives the system into the new phase, the photon field must be excited synchronously in order to maintain the intrinsic dynamical consistency of the system. Therefore, in the energy decomposition, pattern $\lambda_2$ provides compensation, but its direct contribution to reducing the total energy is limited. Instead, it exists more as a photon background.
	
	From Fig.~$\ref{F14}$(d), the pattern $\lambda_3$ is also dominated by atomic components, however, the relative phase between the $\sigma_x$ and $-i\sigma_y$ components in its eigenvector is opposite to that of the pattern $\lambda_1$ (the opposite trends of patterns $\lambda_1$ and $\lambda_3$ in Fig.~$\ref{F14}$(a) also indicate a competitive relationship between them). Therefore, in the energy decomposition, the pattern $\lambda_3$  plays a key role in energy compensation, thereby balancing and stabilizing the new phase. Furthermore, Fig.~$\ref{F14}$ shows that $\lambda_n$ and $u_{n,j}$ exhibit no singular behavior at the critical coupling strength, indicating that phase transition information cannot be obtained from a single pattern alone. The key to the phase transition lies in the competition and balance among the patterns.

We have also simulated the ground state photon number and spin component for each pattern, as shown in Fig.~$\ref{F15}$. Across the entire range of coupling strengths, pattern $\lambda_2$ accounts for nearly the entire weight of the photon number, while its contribution to the spin components is negligible; conversely, patterns $\lambda_1$ and $\lambda_3$ dominate the spin component but contribute little to the photon number. From this, we can conclude that since all photonic components of the system are encapsulated within the pattern $\lambda_2$, any behavior of the photon field, including its mean value (first moment) and fluctuations (second moment) should be governed by the dynamics of the pattern $\lambda_2$.

%

\end{document}